\documentclass[pre,aps,twocolumn,superscriptaddress,nofootinbib]{revtex4-2}
\usepackage{graphicx}
\usepackage{amssymb,latexsym,amsthm,amsmath}    
\usepackage{mathtools}
\usepackage{float}
\usepackage[draft]{changes}
\usepackage{hyperref}
\usepackage{color}
\usepackage{hhline}
\usepackage{colortbl}

\usepackage{ulem} 

\usepackage{xcolor}

\newcommand{\mean}[1]{\left\langle #1 \right\rangle}
\begin{document}
\title{Deterministic and stochastic cooperation transitions in evolutionary games on networks}
\author{Nagi Khalil}
\email{nagi.khalil@urjc.es}
\affiliation{Complex Systems Group \& GISC, Universidad Rey Juan Carlos, Móstoles, 28933 Madrid, Spain}
\author{I. Leyva}
\affiliation{Complex Systems Group \& GISC, Universidad Rey Juan Carlos, Móstoles, 28933 Madrid, Spain}
\affiliation{Center for Biomedical Technology, Universidad Politécnica de Madrid, Pozuelo de Alarcón, 28223 Madrid, Spain }
\author{J.A. Almendral}
\affiliation{Complex Systems Group \& GISC, Universidad Rey Juan Carlos, Móstoles, 28933 Madrid, Spain}
\affiliation{Center for Biomedical Technology, Universidad Politécnica de Madrid, Pozuelo de Alarcón, 28223 Madrid, Spain }

\author{I. Sendi\~na-Nadal}
\affiliation{Complex Systems Group \& GISC, Universidad Rey Juan Carlos, Móstoles, 28933 Madrid, Spain}
\affiliation{Center for Biomedical Technology, Universidad Politécnica de Madrid, Pozuelo de Alarcón, 28223 Madrid, Spain }

\begin{abstract}
Although the cooperative dynamics emerging from a network of interacting players has been exhaustively investigated, it is not yet fully understood when and how network reciprocity drives cooperation transitions. In this work, we investigate the critical behavior of evolutionary social dilemmas on structured populations by using the framework of master equations and Monte Carlo simulations. The developed theory describes the existence of absorbing, quasi-absorbing, and mixed strategy states and the transition nature, continuous or discontinuous, between the states as the parameters of the system change. In particular, when the decision-making process is deterministic, in the limit of zero effective temperature of the Fermi function, we find that the copying probabilities are discontinuous functions of the system's parameters and of the network degrees sequence. This may induce abrupt changes in the final state for any system size, in excellent agreement with the Monte Carlo simulation results. Our analysis also reveals the existence of continuous and discontinuous phase transitions for large systems as the temperature increases, which is explained in the mean-field approximation. Interestingly, for some game parameters, we find optimal "social temperatures" maximizing/minimizing the cooperation frequency/density.
\end{abstract}

\maketitle

\section{Introduction}
Cooperation and defection are both ubiquitous behaviors in natural societies, including indeed those of humans. While defectors usually receive the highest benefits when acting selfishly, cooperators help others in an altruistic way at their own cost and, based on the “survival of the fittest” principle, defection should prevail against cooperation. Yet Nature provides us with numerous examples where cooperative interactions among agents (be either humans, animals, microorganisms, or genes) are at the origin of more complex and functional systems \cite{hamilton_wd_an63,szathmary_jtb97,dugatkin_bp02,doebeli_el05}. 

Understanding the mechanisms driving the evolution of cooperation within a population is  at the core of the Evolutionary Game Theory \cite{axelrod_s81,nowak_s04}. Under this mathematical framework, social dilemmas are modeled as games among agents whose strategies are allowed to spread within the population according to their payoffs through a replicator dynamics \cite{hofbauer_98}. One of the mechanisms known to favor cooperation is the  reciprocity induced by the spatial distribution of the players as shown by Nowak and May \cite{nowak_n92b}. When interactions are no longer well-mixed and players are distributed in a spatial/topological structure, cooperators can cluster together and might survive surrounded by defectors, changing the mean-field equilibrium panorama of many games \cite{hauert_n04,szabo_pr07,perc_review2013,klemm2020altruism}. 

Since that seminal work by Nowak and May \cite{nowak_n92b} and with the rapid development of the complex networks field in the last few decades \cite{boccaletti_pr06,estrada2012structure,boccaletti_pr14}, a lot of research has been focused on the role of the underlying network topology in the emergence of cooperation, including aspects like network heterogeneity both theoretically \cite{santos_prl05,gomez-gardenes_prl07} and experimentally \cite{grujic_pone10,gracia-lazaro_pnas12,rand_pnas11, suri_pone11}, the presence of a layered structure describing the different types of social relationships \cite{gomez-gardenes_srep12,matamalas_srep15}  and degree correlations among layers \cite{wang_pre14a}, or game refinements by introducing topology dependent payoffs \cite{sinha_epjb21} and the influence of an update rule and connectivity on the outcome dynamics of structured populations \cite{roca_plr09,raducha_sr22}.

From a Statistical Physics perspective \cite{hauert_ajp05,perc_pr17}, several attempts have been made to provide rules that predict transitions to collective states of cooperation at critical points, involving the structure connectivity and the game parameters. For example, Ohtsuki et al. \cite{ohtsuki_n06} using mean-field and pair approximations derived a simple condition stating that the ratio of benefit to cost of the altruistic act 
has to exceed the mean degree 
to favor cooperation. Konno \cite{konno_jtb11}, however, suggests that what really matters is the mean degree of the nearest neighbors. Recently, Zhuk et al.\cite{Zhuk2021} showed that the unique sequence of degrees in a network can be used to predict for which game parameters major shifts in the level of cooperation can be expected; this includes phase transitions from absorbing to mixed strategy phases, characterized by agents switching intermittently between cooperation and defection. Using finite-size scaling, Menon et al. \cite{menon_fp18} investigated the different phase transitions between those collective states and found critical exponents dependent on the connection topology. Phase transitions in evolutionary cooperation induced by lattice reciprocity have been also investigated using the standard Statistical Mechanics of macroscopic systems, showing that the onset of the phase transition cannot be captured by a purely mean-field approach \cite{floria_pre09,flores_arxiv22}.

Indeed, due to the intrinsic complexity of games on graphs, their analytical treatment is a  challenging task \cite{gleeson_prl11,amaral2016stochastic,lee_jcn18,peralta_prr20}. Here we present a general analytical approach based on the master and the Fokker-Planck equations derived for a network of pairwise engaged agents whose reproductive success, in terms of the replication rate of their strategy, depends on the payoff obtained during the interaction. In this work, the resulting payoff depends on the actual agents' strategies through a general matrix of payoffs to account for the full space of two-player social dilemmas with two strategies, cooperation and defection. We generalize the results obtained in \cite{Zhuk2021}, by examining in detail the steady and metastable states and the nature (continuous or discontinuous) of the phase transitions between absorbing, quasi-absorbing, and mixed strategy states using the master and Fokker-Planck equations, for any system size and any effective temperature describing how often a player makes irrational choices.

The work is organized as follows. In Section \ref{sec:model} we define the model (game dynamics, updating rule, and interaction network) and introduce the notation used throughout this work. In Section \ref{sec:theory} we study the most general master equation describing the state of the system and identify some steady-state solutions and discontinuity points depending on the parameters of the system. The mean-field case is thoroughly investigated in  Section \ref{sec:meanfield} by means of the corresponding Fokker-Planck equation, which we solve analytically by artificially removing the singularities at the pure absorbing states of the system. Monte-Carlo numerical simulations are provided in Section \ref{sec:validmean} to corroborate the analytical predictions of mean field, both for all-to-all interactions and for more complex interaction networks. Finally, we summarize our results in the Conclusion section.

\section{Model definition}
\label{sec:model}

We consider a population of $\mathcal {N}$ agents playing a $2\times 2$ game, where each agent can adopt a strategy of cooperation (C) or defection (D), that can be changed depending on her performance, her neighbors' performance, and some degree of randomness. The population connectivity is structured in a connected and undirected network represented by the adjacency matrix $\mathcal A$, such that  $\mathcal A_{\mu,\nu}=\mathcal A_{\nu,\mu}=1$ if nodes $\mu$ and $\nu$ are neighbors, while $\mathcal A_{\mu,\nu}=\mathcal A_{\nu,\mu}=0$ otherwise. We denote by $\Sigma$ the set of all nodes and  by $\mathcal{V}_\sigma=\{\nu\in\Sigma \, |\, \mathcal A_{\sigma,\nu}=1\}$ the set of neighbors of a given node $\sigma$.  The number of elements of $\mathcal{V}_\sigma$ is the degree of $\sigma$, $k_\sigma=\sum_{\nu}\mathcal A_{\sigma,\nu}$. Throughout this work, in addition to the complete graph (CG) describing all-to-all interactions, we will consider different graph-structured populations ranging from random regular graphs (RR), Erd\"os–R\'enyi random graphs (ER) \cite{Erdos_1959}, to scale-free networks (SF) using the Barab\'asi-Albert model \cite{Barabasi1999}. 

As any node $\sigma \in\Sigma$ is always occupied by an agent, for our discussion it is useful to use the Boolean variables $c_\sigma$ and $d_\sigma$, indicating if $\sigma$ holds a cooperator or a defector, respectively. Then, it is readily seen that $c_\sigma,\, d_\sigma\in\{0,1\}$, $c_\sigma+d_\sigma=1$, and $c_\sigma\cdot d_\sigma=0$. As a consequence, in order to specify the state $\mathcal S$ of the system at a given time $t$, we only need the set $\mathcal S=\{c_\sigma\, |\,\sigma\in\Sigma\}$.

The dynamics, including Monte Carlos simulations, unfolds in several steps:
\begin{itemize}
\item [(i)] First, the network $\mathcal A$ and an initial state $\mathcal S_0$ are selected.
\item [(ii)] All agents play the game with their neighbors. The resulting payoff of a dyadic interaction is given by the payoff matrix:
\begin{equation}
  \label{eq:pdmatrix}
    {M} =
\begin{array}{c|cc}
 &{\rm C} & {\rm D}\\ \hhline{-|--}
{\rm C} & R  & S\\ \hhline{~|~}
{\rm D} & T & P \\
\end{array}.
\end{equation}
The values $R$, $S$, $T$, and $P$ classically represent the reward for mutual cooperation ($R$), the sucker’s payoff ($S$), the temptation to defect ($T$), and the punishment for mutual defection ($P$). 
 This way, the payoff $g_\sigma$ of an agent at node $\sigma$ depends on the parameters of the matrix $M$, her state, and the state of her neighbors as
\begin{equation}
  \label{eq:gsigma}
  g_\sigma=c_\sigma\sum_{\nu\in \mathcal{V}_\sigma}(Rc_\nu+Sd_\nu)+d_\sigma\sum_{\nu\in \mathcal{V}_\sigma}(Tc_\nu+Pd_\nu).
\end{equation}
\item[(iii)] After the play, an agent at $\sigma$ and one of her neighbors at $\nu$ are selected at random. The former copies the strategy of the latter with a probability
\begin{equation}
  \label{eq:copprob}
  p_{\sigma,\nu}=\frac{1}{1+\exp\left(\frac{-\Delta g_{\sigma,\nu}}{\theta }\right)},
\end{equation}
where $\theta$ is a non-negative parameter playing the role of an effective temperature (tuning the probability of an irrational choice) and
\begin{equation}
  \label{eq:deltagas}
 \Delta g_{\sigma,\nu}=\frac{g_{\nu}-g_\sigma}{T\max(k_\sigma,k_\nu)}
\end{equation}
is a normalized payoff difference.  
\item[(iv)] The time $t$ and the state $\mathcal S$ of the system are updated: $t\to t+t_0$, $\mathcal S\to \mathcal S'$, where $t_0$ is an arbitrary unit of time. 
\item[(v)] The steps (ii) to (iv) are repeated a desired number of times. 
\end{itemize}

For zero effective temperature ($\theta=0$) the copying mechanism is (almost) deterministic: if $\Delta g_{\sigma,\nu}>0$ then node $\sigma$ always copies the strategy of node $\nu$ ($p_{\sigma,\nu}=1)$, while nothing changes when $\Delta g_{\sigma,\nu}<0$ ($p_{\sigma,\nu}=0$). In the tie case  $\Delta g_{\sigma,\nu}=0$, the copying probability is $p_{\sigma,\nu}=\frac12$. In this case ($\theta=0$) and for a very large and well-mixed population, four different categories of games have been extensively studied as a function of the parameters $R$, $S$, $T$, and $P$: Harmony, Snowdrift, Stag Hunt, and Prisoner's Dilemma. The Harmony game represents a category of games satisfying $R>S>P$ and $R>T>P$ where full cooperation is the only possible stable outcome in a population \cite{licht99}, while in the Prisoner's Dilemma, $T>R>P>S$, the evolutionary stable strategy is a whole population of defectors \cite{axelrod_jcr80}. The other two categories represent respectively the classes of anti-coordination and coordination games. In the Snowdrift game \cite{sugden05}, $T>R>S>P$, full defection and cooperation are unstable and the best response is always doing the opposite of your opponent, giving rise to a mixed strategy state. In the Stag Hunt game \cite{skyrms_04}, $R>T>P>S$, players either always cooperate or always defect.

In the opposite temperature limit, when $\theta\to \infty$  the model reduces to the well-known Voter Model \cite{clifford1973model,holley1975ergodic,castellano2009statistical}. In this case, the dynamics is independent of the payoffs ($p_{\sigma,\nu}\to \frac{1}{2}$), and the nodes blindly copy the state of a randomly chosen neighbor. In our study,  we will consider the effects of small and intermediate values of $\theta$ on the final state of the system. 

\section{Theoretical description} \label{sec:theory}
\subsection{Master equation}
Due to the stochastic character of the dynamics and the initial state $\mathcal S_0$, we consider the probability $\mathcal P(\mathcal S,t)$ of finding the system at state $\mathcal S$ at a given time $t$. The dynamics is Markovian and completely determined by the probability rates of the elementary transitions:
\begin{itemize}
\item a change of a defector at a node $\sigma$ to a cooperator,
\begin{equation}
  c_\sigma=0 \underset{\pi^+_\sigma}{\longrightarrow} c_\sigma=1,
\end{equation}
with a rate $\pi^+_\sigma$,
\item and a change of a cooperator to a defector,
\begin{equation}
  c_\sigma=1 \underset{\pi^-_\sigma}{\longrightarrow} c_\sigma=0,
\end{equation}
with a rate $\pi^-_\sigma$.  
\end{itemize}
Note that the dynamics can be seen as a birth-death process, hence suitable for being analyzed as in previous works \cite{khalil2017nonlocal,klemm2020altruism}. Taking into account the steps (ii) and (iii) of the evolution given in the previous section, the rates can be written as 
\begin{eqnarray}
  && \pi_\sigma^+=\frac{d_\sigma}{\mathcal Nk_\sigma t_0}\sum_{\nu\in \mathcal V_\sigma}c_\nu p_{\sigma,\nu}, \\
  && \pi_\sigma^-=\frac{c_\sigma}{\mathcal Nk_\sigma t_0}\sum_{\nu\in \mathcal V_\sigma}d_\nu p_{\sigma,\nu},
\end{eqnarray}
where $p_{\sigma,\nu}$ is provided by Eq.~\eqref{eq:copprob}. 

The probability $\mathcal P(\mathcal S,t)$ obeys the following master equation
\begin{eqnarray}
  \label{eq:mastereq}
  \nonumber
\partial_t \mathcal P(\mathcal S,t)=\sum_{\sigma\in\Sigma} && \left[(\mathcal E_\sigma^+-1)\pi_\sigma^-\mathcal P(\mathcal S,t)\right. \\ && \left. +(\mathcal E_\sigma^--1)\pi_\sigma^+\mathcal P(\mathcal S,t)\right],
\end{eqnarray}
where $\partial_t\mathcal P(\mathcal S,t)\equiv \frac{1}{t_0}\left[\mathcal P(\mathcal S,t+t_0)-\mathcal P(\mathcal S,t)\right]$ is the discrete time derivative and the new operator $\mathcal E_\sigma^+$ ($\mathcal E_\sigma^-$) acts on any function of the state of the system by increasing (decreasing) the number of cooperators at node $\sigma$ by one. 

The master equation can not be solved analytically in general. Nevertheless, some solutions can be identified and analyzed upon changing the parameters of the system. In particular, we will be concerned with the values of the parameters for which there are major changes in the mean fraction of cooperators   $\mean{\rho}$, defined in terms of the probability function $\mathcal P(\mathcal S,t)$ as
\begin{equation}
  \mean{\rho}=\frac{1}{\mathcal N}\sum_{\mathcal S} \sum_{\sigma\in\Sigma} c_\sigma \mathcal P(\mathcal S,t),
\end{equation}
where the sum $\sum_{\mathcal S}$ is over all states. 

\subsection{Steady, absorbing, and quasi-absorbing states}
\label{sec:states}

We assume that, for any initial state, the system always reaches a steady or metastable state. The steady states are characterized by a probability function $\mathcal P(\mathcal S)$ verifying
\begin{equation}
  \label{eq:masteqst}
  \sum_{\sigma\in\Sigma}(\mathcal E_\sigma^+-1)\pi_\sigma^-\mathcal P(\mathcal S)+(\mathcal E_\sigma^--1)\pi_\sigma^+\mathcal P(\mathcal S)=0
\end{equation}
for all states $\mathcal S$. The system \eqref{eq:masteqst} has an infinite number of solutions, including the absorbing states for which $\pi_\sigma^-=\pi_\sigma^+=0$ for all nodes. It is readily seen that the absorbing states are, for any value of $\theta$, the consensus states: $\{c_\sigma=1\}$ (full cooperation) and $\{c_\sigma=0\}$ (full defection).

In the case of positive effective temperature $\theta>0$, the probability of copying a neighbor's strategy is always positive $p_{\sigma,\nu}>0$. Hence, for finite system size $\mathcal N<\infty$, there is a nonzero probability for the system to reach and get trapped into any of the two consensus states starting from any initial state. As a consequence, the only steady-state solutions to the master equation when $\theta>0$ and $\mathcal N<\infty$ are of the form $P(\mathcal S)=p_c\mathcal{P}_c(\mathcal S)+p_d\mathcal{P}_d(\mathcal S)$, i.e. linear combinations of the probability functions $\mathcal{P}_c$ (full cooperation) and $\mathcal{P}_d$ (full defection) with $p_c$ and $p_d$ representing the probabilities of reaching the cooperation and defection consensus states, respectively. 



\begin{figure}[!ht]
    \centering
    \includegraphics[width=0.5\textwidth]{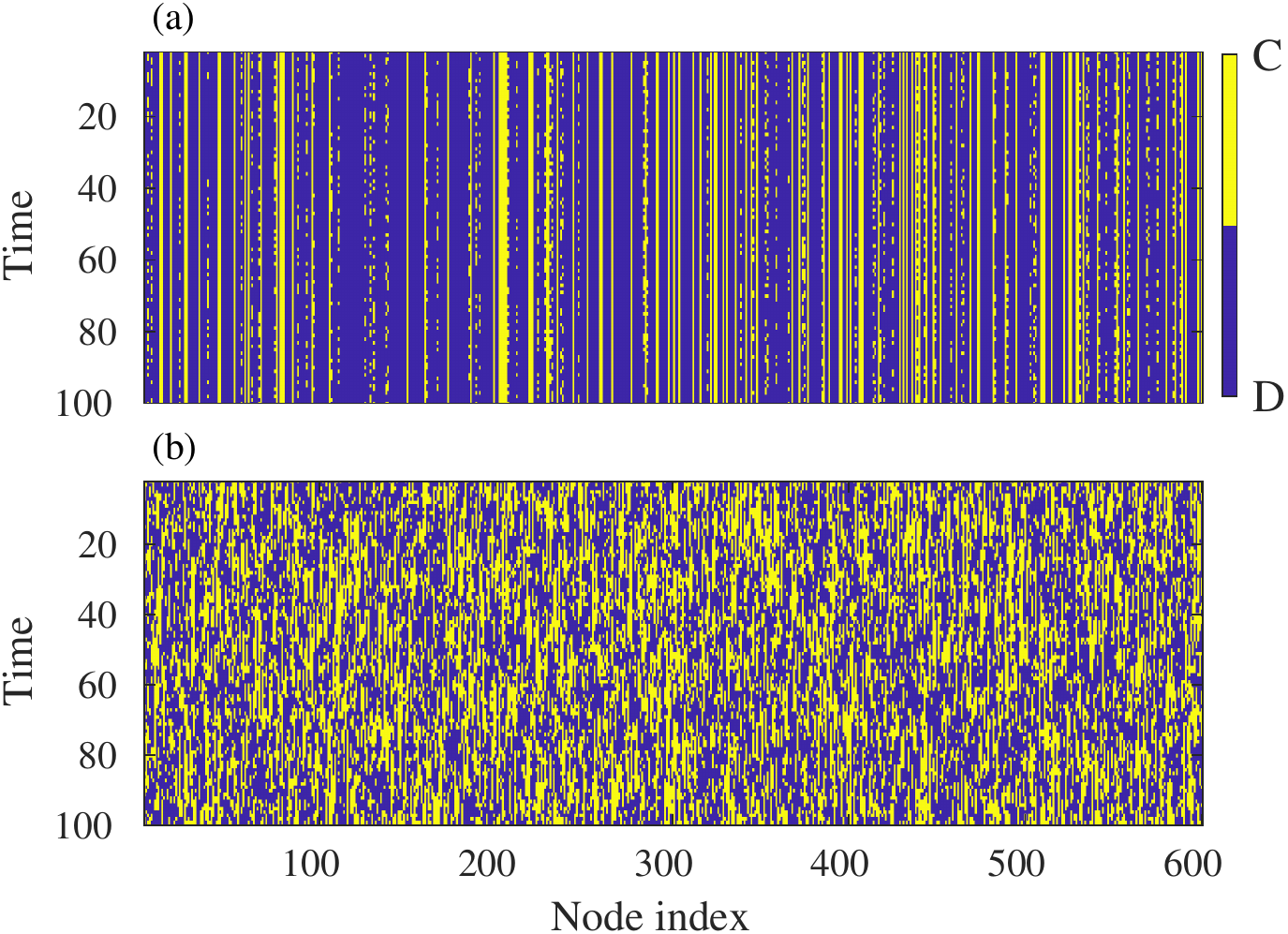}
    \caption{Raster plots for the evolutionary dynamics of a population on a random regular graph for two different game settings displaying quasi-absorbing states (top panel, $S=-0.4,\, T=0.55$) and mixed strategy states (bottom panel, $S=0.4,\, T=1.55$). Blue/yellow colors encode defection/cooperation strategies. The  population size is ${\mathcal N}=600$ and each player is randomly connected to $k=3$ neighbors. The rest of parameters are $R=1$ and $P=\theta=0$.}
    \label{fig:quasi}
\end{figure}

However, the case of zero temperature ($\theta=0$) requires a more careful analysis. Apart from the absorbing states, we find that, depending on the parameters and the network structure, the system can also get trapped to either a set of what can be called {\it quasi-absorbing} states or a set of {\it mixed strategy} states. We show two raster plot examples in Fig.~\ref{fig:quasi}. In the quasi-absorbing states, which appear mainly but not only when $S<0$ and $T<1$, see Fig.~\ref{fig:quasi}(a), connected domains of nodes with frozen strategy are separated by a frontier of oscillating nodes. In addition, the system can also get trapped to a set of mixed strategy states, in which all nodes change their strategy along the evolution. These latter states appear for $S>0$ and $T>1$, as shown in Fig.~\ref{fig:quasi}(b). From a dynamical viewpoint, both the quasi-absorbing and mixed strategy states form absorbing sets of states (once reached, the system has no way to leave them). 

The quasi-absorbing states have been previously studied in a similar model \cite{gomez-gardenes_prl07}, being there the "locally fluctuating strategies". In that work the dynamic rules are different: an agent never copies the strategy of others with smaller payoff, but that of higher payoff with some probability. That is, the rule includes a deterministic part. Hence, we conclude that, for the existence of the quasi-absorbing states, at least some degree of determinism in the update rule is required. 

Mathematically, as already noted, the quasi-absorbing and mixed strategy states form a self-absorbing subset of states, being the absorbing states a special limiting case of the former. Absorbing, quasi-absorbing, and mixed strategy states make the dynamics non-ergodic, i.e. only a subset of states can be explored from a given initial condition. Only with an effective temperature big enough, and disregarding the absorbing states, and we can ensure an ergodic dynamics.

Finally, it is worth stressing that when the size of the system increases and/or for some values of its parameters, the time required to reach the absorbing and/or quasi-absorbing states may grow very fast, both for $\theta=0$ and $\theta>0$. This way, at the relevant physical scales, the system is very often at (macroscopic) metastable states. This already happens in the Voter Model \cite{vazquez2008analytical}, which is the $\theta\gg \mathcal N$ limit of our model. But metastability also occurs in the limit of small effective temperature when, for instance, the system stays close to the quasi-absorbing states. Metastability of mixed strategy states already happens with all-to-all interactions, as we analyze in Sec.~\ref{sec:meanfield}.

\subsection{Transition points}
\label{sec:transitionpoints}


The parameters of the system, together with the initial conditions, determine the evolution and final states of the system through the dependence of the rates $\pi^\pm$ on them. This is very apparent, for instance, when analyzing the steady-state solutions $\mathcal P_{st}$ to the master equation. 
Consider Eq.~\eqref{eq:masteqst} for $\mathcal P_{st}$, which can be written in matrix form as
\begin{equation}
  \mathcal W \vec{\mathcal P}_{st}=\vec{0},
\end{equation}
where $\mathcal W$ represents the $2^\mathcal N\times 2^\mathcal N$ matrix of coefficients and the vector $\vec{\mathcal P}_{st}$ is obtained by evaluating $\mathcal P_{st}$ at all $2^\mathcal N$ possible states. For a finite-size system ($\mathcal N<\infty$) with positive temperature ($\theta>0$) the rates $\pi^\pm$ and, hence, all the components of $\mathcal W$ are smooth functions of the parameters. Therefore, $\vec{\mathcal P}_{st}$ (also $\mathcal P_{st}$) depends continuously on the parameters of the system. However, for $\theta=0$ the copying probability $p_{\sigma,\nu}$ in Eq.~\eqref{eq:copprob} is a strongly discontinuous function of the system parameters (see Appendix A for details). The discontinuities may induce jumps on the steady-state solutions as we change the parameters. Similar arguments can be applied for the metastable states as well. Taking the limit $\theta \to 0$ in the expression \eqref{eq:copprob}, the possible discontinuities can be localized with the condition
\begin{equation}
  \label{eq:deltageq0}
  \Delta g_{\sigma,\nu}=0,
\end{equation}
provided two agents with different strategies are located at $\sigma$ and $\nu$. Using Eqs.~\eqref{eq:gsigma} and \eqref{eq:deltagas} with condition \eqref{eq:deltageq0}, we derive the following conditions
\begin{eqnarray}
  \nonumber && mR+(k_\sigma-m)S=nT+(k_\nu-n)P, \\ && \quad 0\le m\le k_{\sigma} -1, \qquad 1\le n\le k_{\nu}; \label{eq:diccond1}  
\end{eqnarray}
for natural numbers $m$ and $n$, and when there is a non-zero probability of finding a cooperator with degree $k_\sigma$ with at least one defector neighbor with degree $k_\nu$. Note that the static condition \eqref{eq:diccond1}, given by the degree sequence of the network, is restricted by an additional dynamic condition, which induces a eventual dependence on the initial conditions.

\begin{figure}[!ht]
    \begin{flushright}
    \includegraphics[width=0.5\textwidth]{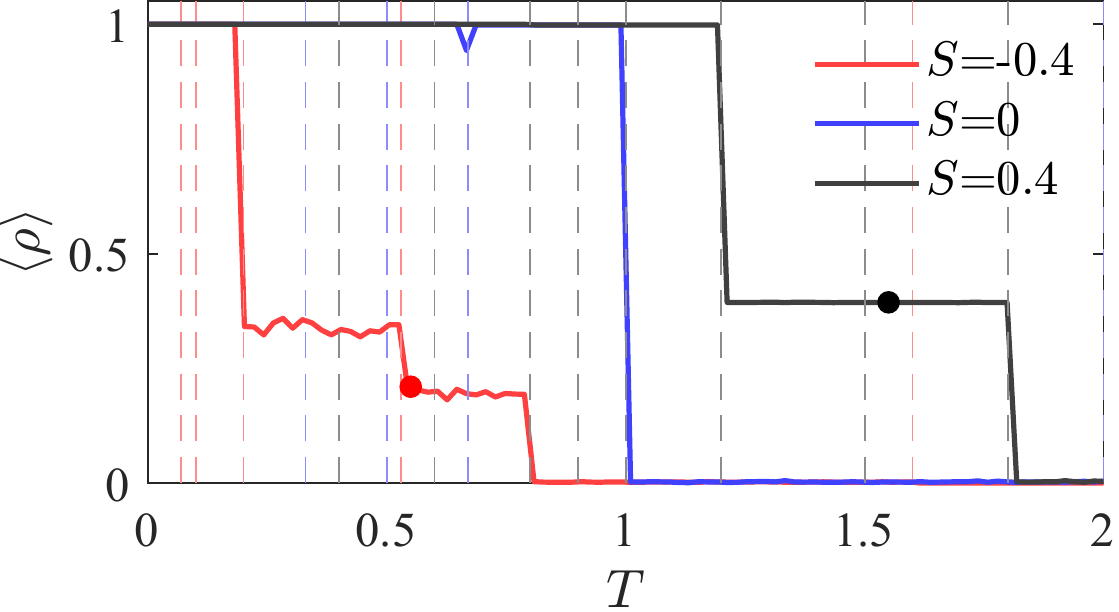}
    \includegraphics[width=0.44\textwidth]{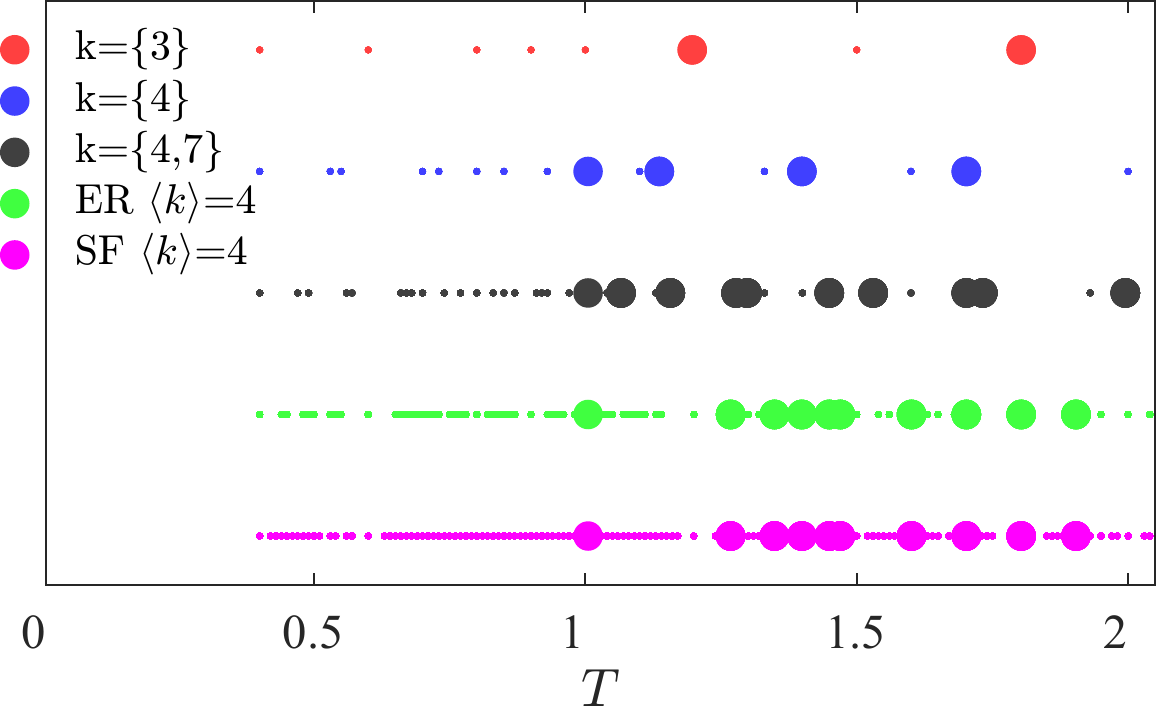}
    \caption{Deterministic cooperation transitions.  (a) Average cooperation observed in Monte Carlo simulations of random regular graphs with ${\mathcal  N}=1000$ nodes and $k=3$ as a function of $T$ for three different values of $S$ (see legend). Vertical dashed lines correspond to the theoretical values of $T$ fulfilling Eq.~\eqref{eq:diccond1} for each one of the values of $S$ used. (b)  Small dots show all the $T$ values solving  Eq.~\eqref{eq:diccond1} for different network degree distributions (see legend) and $S=0.4$. Among these values, those leading to an  actual major shift of the cooperation in our simulations are highlighted with larger dots. In panel (a), the red (black) circle corresponds to the game dynamical settings used in Fig.~\ref{fig:quasi} (a)((b)) panel. See main text for the rest of parameters. 
    \label{fig:jumps}}    
    \end{flushright}        
\end{figure}

 The previous abrupt changes in the distribution function for $\theta$=0 translates into discontinuities in the average cooperation $\mean{\rho}$, as illustrated in Fig.~\ref{fig:jumps}. Along the work, unless otherwise said, all simulations correspond to networks with $\mathcal N=1000$, different topologies, $3 \cdot 10^4$ evolution time steps, payoff parameters $R$=1 and $P$=0, and all the results averaged over at least $100$ independent simulations (realizations). Moreover, all initial (microscopic) states $\mathcal S_0$ have the same number of defectors and cooperators randomly placed at the nodes. In  Fig.~\ref{fig:jumps}(a) we show the results for random regular networks with all players having $k=3$ neighbors (therefore $0\le m\le 2$ and $1\le n\le 3$), as a function of $T$. Equation \eqref{eq:diccond1} predicts that discontinuities may only occur at particular values of $T$, shown as vertical dashed lines. Indeed, when we chose values of $S$ exploring the different possible games in the parameter space, we observe that, in all cases, whenever there is a jump in the average cooperation, it precisely coincides with one of the $T$ values fulfilling Eq.~\eqref{eq:diccond1}. The accuracy of this equation, which relates topological features and game dynamics at the microscopic level, holds for different network degree distributions as shown in the bottom panel of Fig.~\ref{fig:jumps}. As the local connectivity patterns become more complex with more degrees present in the distribution (from random regular to ER and SF configurations), the number of combinations matching the condition given by Eq.~\eqref{eq:diccond1} increases (small dots) but not all of them give rise to an actual abrupt change in the cooperation frequency (big dots). Once again, we notice that this is consistent with the fact that, in order for the (static) condition \eqref{eq:diccond1} to be fulfilled, we need the additional (dynamic) condition of having a cooperator with degree $k_\sigma$ and a defector neighbor with degree $k_\nu$. 
 
Notice that even if the macroscopic evolution of the averaged cooperation $\langle \rho \rangle$ in Fig.~\ref{fig:jumps}(a) shows similar qualitative characteristics for all the values of $S$, with dramatic changes separated by long stable plateaus, the microscopic states involved are very different, as discussed in the previous section. For $S<0$ (red curve) the discontinuous transitions occurs between absorbing and quasi-absorbing states, as in Fig.~\ref{fig:quasi}(a),  while in the case $S>0$ (black curve) the cooperation absorbing state yields to mixed strategy states, see Fig.~\ref{fig:quasi}(b), and then to quasi-absorbing states with a very small level of cooperation. Finally, for $S=0$ (blue) curve, the transition is between the cooperation state and a set of quasi-absorbing states very close to the defection consensus.  

 \begin{figure*}[!ht]
    \centering
    \includegraphics[width=0.8\textwidth]{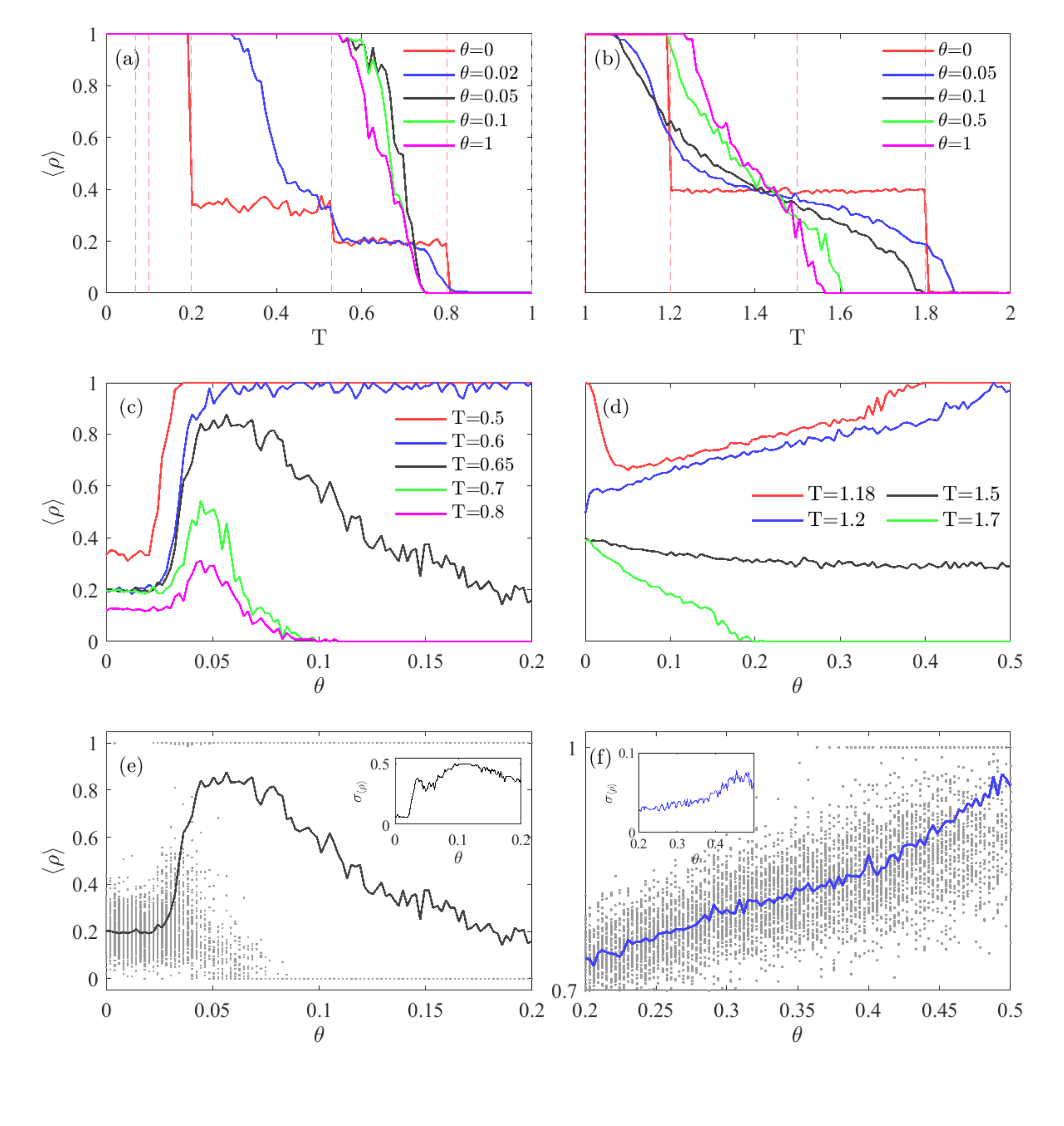}
    \caption{Stochastic cooperation transitions in a random regular network with $\mathcal N=1000$ nodes and $k=3$. (a)-(b) Average cooperation as a function of $T$ for several values of the Fermi temperature  $\theta$ and (a) $S=-0.4$ and (b) $S=0.4$. In both panels, vertical dashed lines correspond to the $T$ values predicted by Eq.~\eqref{eq:diccond1} using the constant $S$ value and the rest of parameters. (c)-(d) Scatter plots showing the average cooperation as a function of $\theta$ resulting from  different Monte Carlo simulations for (f) $T=0.65$ and $S=-0.4$ [black curve in panel (c)] and (g) $T=1.2$ and $S=0.4$ [blue curve in panel (d)]. Insets show the corresponding standard deviation. The rest of parameters in all panels are $R=1$ and $P=0$.}
    \label{fig:noise}
\end{figure*}
 
As long as $\theta>0$ is small enough, we also expect important changes in the behavior of the system when conditions \eqref{eq:diccond1} hold. As shown in Fig.~\ref{fig:noise}(a)-(b), the presence of an stochastic component in the choice of strategy ($\theta>0$) promotes a smoother evolution of the average cooperation as a function of $T$, while the steepest changes still occur near a predicted transition point given by Eq.~\eqref{eq:diccond1}. 

Nevertheless, as the temperature increases, the behavior critically depends on whether the system is in  quasi-absorbing states, Fig.~\ref{fig:noise}(c), or passes through mixed strategy states, Fig.~\ref{fig:noise}(d). In the former case, for some values of the parameters not close to the abrupt transitions, and for very small $\theta$, quasi-absorbing states are "stable" up to a critical temperature $\theta^c$ [$\theta^c\sim 0.02$ in Fig.~\ref{fig:noise}(c)]. Beyond $\theta^c$, quasi-absorbing states are "destroyed" in favor of one of the two possible absorbing states, pure cooperation or pure defection, being the probability of the selection dependent on the values of $T$ and $\theta$. This is plain in the scatter plot in Fig.~\ref{fig:noise}(e), in which each dot is the cooperation density from a single Monte Carlo simulation and, for $\theta>\theta_c$, they are placed at either 1 (full cooperation) or 0 (full defection). These numerical results suggest the existence of a discontinuous thermodynamic-like transition (for $\mathcal N\gg 1$) separating two distinct phases. For $\theta<\theta_c$ the system is in a (macroscopic) mesoscopic state around the quasi-absorbing states (identified at $\theta=0$); while for $\theta>\theta_c$ the system reaches either full cooperation or defection with a temperature dependent probability. This picture is also sustained by the presence of a peak in the dispersion of dots (standard deviation) around $\theta_c$, as shown in inset of Fig.~\ref{fig:noise}(e).
It is also remarkable the resonant behavior of the probability of reaching consensus with the temperature $\theta$ for some values of $T$. As the temperature increases, the average cooperation is promoted up to a maximum; beyond this peak, higher values of $\theta$ favors defection instead.    

On the other hand, the mixed strategy states are more robust in a broad range of temperatures and we observe a monotonous behavior with increasing/decreasing or constant values of the average fraction of cooperators as the temperature $\theta$ rises, Fig.~\ref{fig:noise}(d). However, a non-monotonic behavior of the cooperation density as a function of $\theta$ is also observed for some values of $T$ [see for example $T=1.18$ in Fig.~\ref{fig:noise}(d)]. In this case, the system loses the pure cooperation state as the temperature increases, reaching a minimum of cooperation for intermediate temperatures. Here as well, the simulations suggest that the temperature acts as a control parameter driving the system through a discontinuous phase transition, but this time between mixed strategy states and full cooperation, as shown in the scatter plot of Fig.~\ref{fig:meanfield}(f) for $T=1.2$. 

Finally, it is worth noting that, on the one hand, the abrupt behavior of the system for $\theta\ge 0$ given by conditions \eqref{eq:diccond1} does not require to take the thermodynamic limit ($\mathcal N\to \infty$), it occurs for any system size. On the other hand, the first-order-like transitions suggested by Figs.~\ref{fig:meanfield}(e)-(f) do require the thermodynamic limit. 

\section{Mean field}\label{sec:meanfield}
In the case of having an all-to-all connectivity, all agents can be regarded as physically equivalent and, thus, the system can be fully described with the probability function $\mathcal P(\rho,t)$ of finding a fraction of cooperators $\rho$ at a time $t$. This probability is defined as
\begin{equation}\label{eq:mfP}
  \mathcal P(\rho,t)={\sum}^\rho\mathcal P(\mathcal S,t),
\end{equation}
where ${\sum}^\rho$ stands for the sum over all states with the same fraction of cooperators $\rho$. Using that  $  \mathcal P(\mathcal S,t)=\mathcal P(\mathcal S',t)$ for any two states $\mathcal S$ and $\mathcal S'$ with the same $\rho$, Eq.~\eqref{eq:mfP} can be expressed as 
\begin{equation}
  \label{eq:pmeanf}
  \mathcal P(\rho,t)=\frac{\mathcal N!}{(\mathcal N \rho)!(\mathcal N(1-\rho))!}\mathcal P(\mathcal S,t),
\end{equation}
for any state $\mathcal S$ with a fraction of cooperators $\rho$.

\subsection{The master equation}

A master equation for the $\mathcal P(\rho,t)$ can be obtained using Eq.~\eqref{eq:pmeanf} and by summing both sides of Eq.~\eqref{eq:mastereq} over all states $\mathcal S$ with a fraction of cooperators $\rho$,
\begin{equation}
  \label{eq:mastereqrho}
  \partial_t \mathcal P(\rho,t)=(\mathcal E^+-1)\pi^-\mathcal P(\rho,t)+(\mathcal E^--1)\pi^+\mathcal P(\rho,t),
\end{equation}
where $\mathcal E^+$ ($\mathcal E^-$) increases (decreases) the argument of any function of $\rho$ by ${1}/{\mathcal N}$, and the new rates $\pi^\pm$ read
\begin{eqnarray}\label{eq:mfpimas}
  && \pi^+=\frac{\mathcal N}{(\mathcal N -1)t_0}\rho(1-\rho) p^+, \\\label{eq:mfpimenos}
  && \pi^-=\frac{\mathcal N}{(\mathcal N -1)t_0}\rho(1-\rho) p^-,
\end{eqnarray}
with
\begin{equation}
  \label{eq:ppm}
p^\pm=\left[1+\exp\left(\frac{-\Delta g^\pm}{\theta}\right)\right]^{-1}
\end{equation}
and
\begin{eqnarray}
  \nonumber 
  \Delta g^\pm= && \mp \frac{1}{T(\mathcal N-1)}\left\{T\mathcal N\rho+P[\mathcal N(1-\rho)-1] \right. \\
                && \left. -[R(\mathcal N \rho-1)+S\mathcal N(1-\rho)] \right\}.
                   \label{eq:gpm}
\end{eqnarray}

For the typical values of $\mathcal N$ (not necessarily in the thermodynamic limit), Eqs.\eqref{eq:mfpimas},\eqref{eq:mfpimenos} and \eqref{eq:gpm} can be very accurately approximated by:
\begin{eqnarray}
  \label{eq:pipm}
  && \pi^\pm \simeq \frac{1}{t_0}\rho(1-\rho){p^\pm}, \\
  && \frac{\Delta g^\pm}{\theta} \simeq \mp\left[t_\theta \rho+s_\theta(1-\rho)\right],
\end{eqnarray}
where the parameters $t_\theta$ and $s_\theta$ are defined as
\begin{eqnarray}
  \label{eq:ttheta}
  && t_\theta=\frac{T-R}{T\theta}, \\
  \label{eq:stheta}
  && s_\theta=\frac{P-S}{T\theta}.
\end{eqnarray}
Notice that, under these simplifications, all the system dependency on the payoffs parameters and the effective temperature $\theta$ occurs through these two new parameters $t_\theta$ and $s_\theta$. In particular, this means that, with high accuracy, a change in the effective temperature (beyond $\theta=0$) is equivalent to keeping the temperature fixed and appropriately changing the game's parameters. 

In the continuum limit, $\partial_t$ is the time derivative and, for $\mathcal N\gg 1$, $\rho\in[0,1]$ becomes a continuum variable. Then, expanding the right-hand side of Eq.~\eqref{eq:mastereqrho} up to order $\left(\frac{2}{\mathcal N}\right)^2$ we obtain the following Fokker-Planck equation:

\begin{eqnarray}
  \nonumber
  \partial_t\mathcal P(\rho,t)&\simeq &-\frac{1}{\mathcal N}\partial_\rho\left[\left(\pi^+-\pi^-\right)\mathcal P(\rho,t)\right]\\ && +\frac{1}{2\mathcal N^2}\partial_\rho^2\left[\left(\pi^++\pi^-\right)\mathcal P(\rho,t)\right], \label{eq:fokker}
\end{eqnarray}
which includes two contributions to the time evolution of $\mathcal P(\rho,t)$. The first one on the right-hand side of the equation is the {\it drift term}, which is proportional to
\begin{equation}
  \pi^+-\pi^-=\rho(1-\rho)(p^+-p^-)
\end{equation}
and vanishes for $\rho=0, 1$ (the absorbing states) and for $p^+=p^-$. This latter condition gives rise to  
\begin{equation}
  \label{eq:rho0}
  \rho\simeq \rho_0\equiv \frac{s_\theta}{s_\theta-t_\theta}=1+\frac{t_\theta}{s_\theta-t_\theta}.
\end{equation}
The second contribution is the  {\it diffusion term} which also vanishes for $\rho=0$ and $\rho=1$ as expected, but it is positive in the interval $\rho\in(0,1)$. In fact,  we have $p^++p^-=1$ and the diffusion term is proportional to $\pi^++\pi^-=\rho(1-\rho)$, which has the same form as in the Voter Model.

Equation~\eqref{eq:fokker} provides a good estimation of $\mathcal P$, including finite-size effects. In particular, we can directly identify three regimes: (a) When $\pi^+-\pi^-$ and $\pi^++\pi^-$ are of the same order (with is the case when $t_\theta,s_\theta\gg 1/\mathcal N$), 
the drift term acts on a time scale $t_1 \sim\mathcal N t_0$ while the diffusion term acts on $t_2\sim \mathcal N^2 t_0$. That is, for $\mathcal N\to \infty$ the drift term is dominant. As we will explicitly show, in this case the drift term may create metastable states with a lifetime of the order of $t_2$, as already mentioned in Sec.~\ref{sec:states}. We will also show that under some conditions, the dynamics is given by the replicator equation. (b) When, but not only, the effective temperature is big enough $\theta \sim \mathcal N$ (which is the case when $t_\theta,s_\theta\sim 1/\mathcal N$), both drift and diffusion terms are of the same order and evolve in the same time scale $t_2\sim\mathcal N^2t_0$. (c) Finally, for $t_\theta=s_\theta=0$, or for extremely large effective temperature $\theta \gg \mathcal N$, the dominant term is the diffusion one and the model becomes the Voter Model. 

\subsection{Effective potential}

Due to the nonlinear dependence of $p^\pm$ on $\rho$ with absorbing states, it is difficult to solve the Fokker-Planck equation analytically (the steady-state solutions are linear combinations of delta functions at $\rho=0,1$). Nonetheless, we can gain some relevant information if we artificially remove the singularities at $\rho=0,1$ as:
\begin{eqnarray}
  && \pi^+-\pi^-\to [\rho(1-\rho)+\kappa](p^+-p^-) , \\
  && \pi^++\pi^-\to [\rho(1-\rho)+\kappa],
\end{eqnarray}
with $\kappa>0$ a small parameter. This parameter makes the system slightly away from the absorbing states, as if we retain one agent of each kind (then $\kappa\sim 1/\mathcal N$). 
With this regularization, we can focus on the steady-state solutions of Eq.~\eqref{eq:fokker}, characterized by a zero probability flux:
\begin{equation}
  -\left(\pi^+-\pi^-\right)\mathcal P(\rho)+\frac{1}{2 \mathcal N}\partial_\rho\left[\left(\pi^++\pi^-\right)\mathcal P(\rho)\right]=0.
\end{equation}
The solution to this equation can be written as 
\begin{equation}
  \label{eq:prhost}
  \mathcal P(\rho)=\mathcal C e^{-2 \mathcal N V(\rho)},
\end{equation}
with $\mathcal C$ a normalization constant and $V(\rho)$ an effective potential given by
\begin{eqnarray}
  \nonumber  V(\rho)=&& {\frac{2}{t_\theta-s_\theta}}\ln\left\{\cosh\left[\frac{t_\theta\rho+s_\theta(1-\rho)}{2}\right]\right\}  \\ && + {\frac{1}{2\mathcal N}}\ln{\left[\rho(1-\rho)+\kappa\right]}, \label{eq:veff}
\end{eqnarray}
valid for $t_\theta\ne s_\theta$.

When $\theta\ll\mathcal N$, in the regime (a) discussed above, we can locate one extreme of $V(\rho)$ at
\begin{equation}
  \label{eq:rhoc}
  \rho_m\simeq \rho_0-\frac{1}{\mathcal N}\frac{t_\theta+s_\theta}{t_\theta s_\theta},
\end{equation}
where $\rho_0$ is given by Eq.~\eqref{eq:rho0}. The previous expression is accurate up to order $\mathcal O(1/\mathcal N^2)$. 
The other two possible extremes can appear closer to the absorbing states ($\rho=0,1$). With the same accuracy, when $\rho_0$ is far from $0$ and $1$, the second derivative of $V$ at $\rho_m$ is
\begin{equation}
  \label{eq:vpp}
  V''(\rho)\simeq t_\theta-s_\theta.
\end{equation}
Hence, $\rho_m$ is a minimum of the potential for $t_\theta> s_\theta$. 
In this case, Eq.~\eqref{eq:rhoc} gives the cooperation density of a state of coexistence of strategies, that is, a mixture of cooperators and defectors with $\rho_m\in(0,1)$. Moreover, from a dynamic point of view, the system is in a metastable state that decays to the absorbing states on a time scale of the order of $t_2\sim \mathcal N^2 t_0$. Assuming $\rho_m\simeq \rho_0$, the minimum $\rho_m$ is in $(0,1)$ only when $P<S$ and $T>R$. For the typical values $R=1$ and $P=0$ often chosen in the literature, in the $T-S$ plane, the game compatible with a minimum $\rho_m \in (0,1)$ is the Snowdrift [points E and F in Fig.\ref{fig:meanfield}(a)]. In the Stag Hunt game, $\rho_m\in(0,1)$ represents a maximum of the potential, and hence a repulsive point [see the A, B, and C points in Fig.\ref{fig:meanfield}(a) and the corresponding curves for the potential in Fig.~\ref{fig:meanfield}(d)].

In the other two regimes, (b) and (c) discussed above, when the effective temperature is large enough, the previous expressions, Eq.~\eqref{eq:rhoc} for $\rho_m$ and Eq.~\eqref{eq:vpp} for  the second derivative of the potential, are no longer valid. The respective new expressions have to be obtained directly by numerically solving $V'=0$, $V''=0$. However, some relevant cases can be addressed analytically, see Appendix \ref{appen:1}. 

\subsection{The replicator equation}

Our mean-field description using the Fokker-Planck equation is more general than the one based on the mean fraction of cooperators
\begin{equation}
  \mean{\rho}(t)=\int d\rho\, \rho \mathcal P(\rho,t),
\end{equation}
which is typically assumed to obey the replicator equation. Therefore, it is interesting here to discuss under what conditions the replicator equation emerges from the Fokker-Planck equation.  

Multiplying Eq.~\eqref{eq:fokker} by $\rho$ and integrating over all values of $\rho$, we obtain:
\begin{equation}
  \partial_t\mean{\rho}=\frac{1}{\mathcal N t_0}\mean{\pi^+-\pi^-}-\frac{1}{2\mathcal N^2 t_0}\left[(\pi^++\pi^-)\mathcal P\right]_0^1.
\end{equation}
When the effective temperature is small ($\theta\ll \mathcal N$) and  the system is not close to the boundaries (such that $\kappa \mathcal P(\rho=0,1)\ll \mathcal N$), we can disregard the diffusion contribution. We then get
\begin{equation}
  \partial_t\mean{\rho}\simeq -\frac{1}{\mathcal N t_0}\mean{\rho(1-\rho)\tanh\left[\frac{t_\theta\rho+s_\theta(1-\rho)}{2}\right]},
\end{equation}
where we have used Eq.~\eqref{eq:pipm} to replace $\pi^\pm$ and removed $\kappa$. Notice that the previous equation is not closed in the sense that it involves moments of $\mathcal P(\rho,t)$ beyond $\mean{\rho}$ that are unknown. Hence, further simplifications are needed. 

When $\mathcal P(\rho,t)$ accumulates around $\mean{\rho}$ then 
\begin{equation}
\partial_t\mean{\rho}\simeq \frac{- \mean{\rho}(1-\mean{\rho})}{\mathcal N t_0}\tanh\frac{t_\theta\mean{\rho}+s_\theta(1-\mean{\rho})}{2}.
\end{equation}
Finally, for $\frac{t_\theta\mean{\rho}+s_\theta(1-\mean{\rho})}{2}\ll 1$, we get the replicator equation
\begin{equation}
  \partial_t\mean{\rho}\simeq -\frac{1}{2\mathcal N t_0} \mean{\rho}(1-\mean{\rho})\left[t_\theta\mean{\rho}+s_\theta(1-\mean{\rho})\right].
\end{equation}

\section{Validity of Mean Field}
\label{sec:validmean}

\subsection{All-to-all interactions}


\begin{figure*}[t!]
    \centering
    \includegraphics[width=1\textwidth]{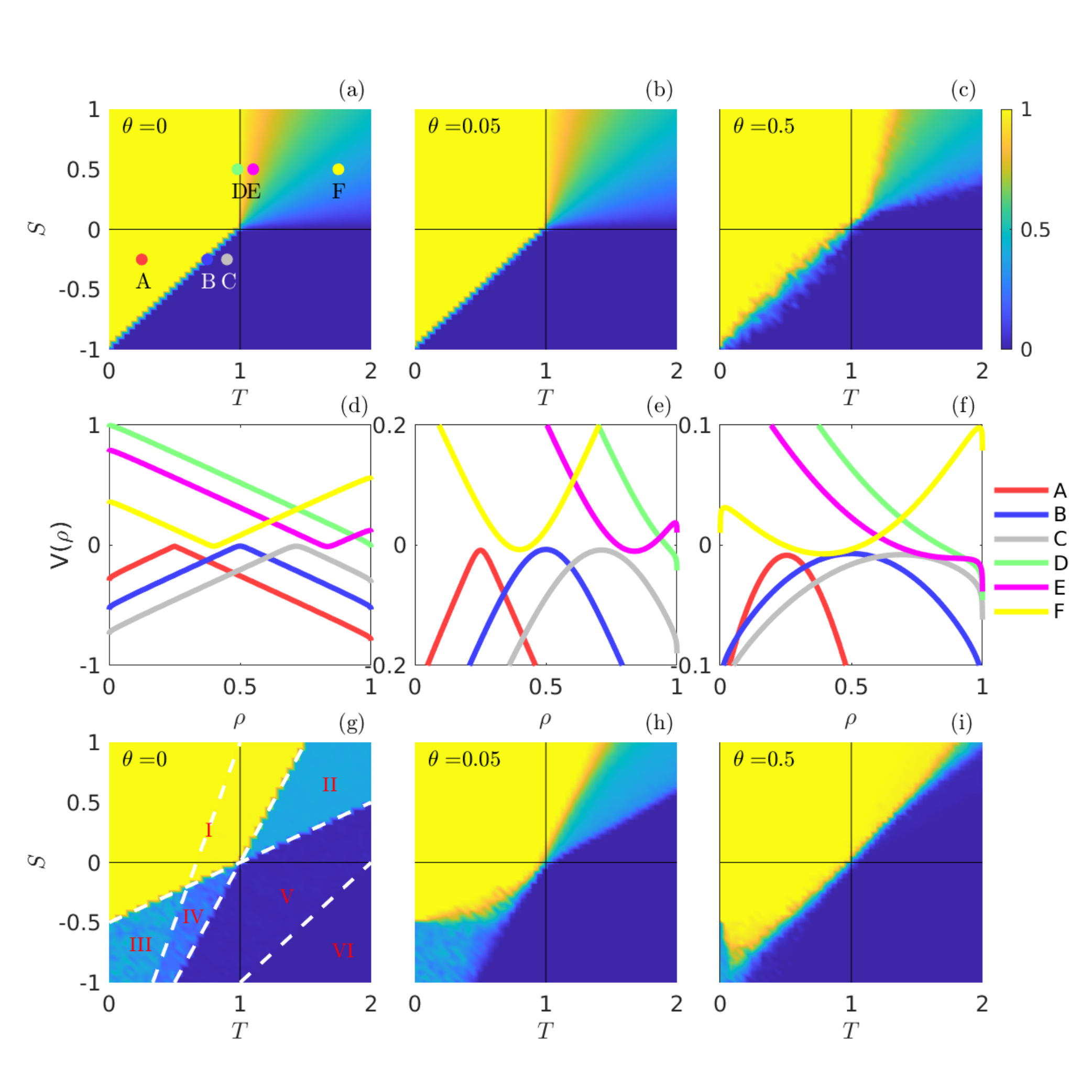}
    \caption{Mean field and the effective potential in the limit $\theta\ll \mathcal N$. (a)-(c) Cooperation density in the $(T,S)$ plane for a complete graph  and effective temperatures (a) $\theta=0$, (b) $\theta=0.05$, and (c) $\theta=0.5$. (d)-(f) Effective potential functions $V(\rho)$ for the parameter settings marked with letters from A to F in panel (a). Each panel corresponds to the same effective temperature as in the upper panel, and the color code is the same as for the symbols labeled in panel (a). Notice the different vertical scales. (g)-(i) Cooperation density in the $(T,S)$ plane for a regular random graph with $k=3$ and the same temperatures as in the top and middle panels. The white dashed lines in panel (g) are solutions to Eq.~\eqref{eq:diccond1} for the $(n,m)\leq k$ pairs (1,1),(1,2),(2,2),(3,2). Initial cooperation level is set to $\%$50 and networks have $\mathcal N=100$ nodes.}
    \label{fig:meanfield}
 \end{figure*}
 
 We validate the mean-field approach performing numerical simulations of agents all-to-all connected and in the regime of small effective temperature. The (a,b,c) panels of Fig.\ref{fig:meanfield} show the cooperation density in the $S-T$ plane for different values of $\theta$. For $\theta=0$, panel (a), the simulations essentially coincide with those predicted by the replicator equation: in the Harmony game ($S>0$ and $T<1$) the system always reaches the cooperation absorbing state, whereas in the Prisoner's Dilemma ($S<0$ and $T>1$) complete defection is the final state. 
 In the Stag Hunt game ($-1<S<0<T<1$) the final outcome is also of complete cooperation and/or defection, depending on the game's parameters and the initial conditions, while for the Snowdrift game ($0<S<1$ and $1<T<2$) a mixture of strategies is found. 
 
 The different observed behaviors are consistent with the shape of the effective potential, $V(\rho)$ in Eq.~\eqref{eq:veff}, as shown in Fig.~\ref{fig:meanfield}(d) for several representative points (A-F) in the $T-S$ plane. In all cases, the steady state of the system is given by the minimum of the potential  closest to the initial  fraction of cooperators ($50\%$ in all the simulations shown in this work). 
 For the region $0<S<1<T<2$ (Snowdrift game), the effective potential has three local minima: two at $\rho=0,1$ and another one at $\rho=\rho_m\in(0,1)$, the latter being dominant for any initial condition different from the consensus states. In this region, the minimum $\rho_m\simeq \frac{S}{S+T-1}\in(0,1)$ is a continuous function of $T$ and $S$, meaning that the cooperation density changes smoothly in this game. At line $T=1,\, S>0$, where the Snowdrift game becomes the Harmony game, $\rho_m\to 1$ and the cooperation is always reached for any initial condition different from complete defection, showing that the interplay between the two games is smooth. Analogously, at the boundary line, $T>1,\, S=0$, where the Snowdrift game turns into the Prisoner's Dilemma game, $\rho_m\to 0$ and both games face the same behavior and, again, no abrupt change is observed. Only in the case of the Stag Hunt game abrupt changes can be observed. If we set the initial value of $\rho$, the final fraction of cooperators (zero or one) depends discontinuously on $S$ and $T$. 
 Alternatively, if we set the parameters $T$ and $S$, the final state depends discontinuously on the initial fraction of cooperators. Finally, when the effective potential is monotonic [not shown in Fig.~\ref{fig:meanfield}(d)] in $\rho\in(0,1)$ (excluding regions near $\rho=0,1$), a slight variation of $T$ and $S$ does not change the final absorbing states of the system, either be complete cooperation or complete defection. This is the case of the Harmony and Prisoner's Dilemma games.

Notice that the apparent coexistence along the line $S=T-1$ in the Stag Hunt game, see Fig.~\ref{fig:meanfield}(a), occurs only when the initial number of cooperators and defectors is the same. For other initial conditions, the region moves to another location. In any case, it corresponds to a situation of an "artificial" coexistence of strategies: the system does not show a mixure of strategies but has a nonzero probability to reach any of the two consensus states. This is in agreement with our (stochastic) theoretical description, and can not be explained using the (deterministic) replicator equation. More precisely, for a given initial fraction of cooperators $\rho=\rho_m$ that cancels $\Delta g^\pm$ in Eq.~\eqref{eq:gpm}, the rates of increasing and decreasing $\rho$ are the same and different from zero (the probability $p^\pm$ in Eq.~\eqref{eq:ppm} is equal to $1/2$). For $\rho\ne \rho_m$, the dynamics is deterministic in the sense that the only possibility is either increasing or decreasing $\rho$, although the time it takes is stochastic. The mean-field description provided by the Fokker-Planck equation is less precise: when an initial fraction of cooperators is close to (not necessarily at) the unique maximum of the effective potential (but not necessarily at $\rho=\rho_m)$, a fluctuation enables the system to cross it and reach either of the two absorbing states. Only in the limit of infinite system size the width of this region tends to zero. 

So far, we have only considered the case $\theta=0$. As long as $\theta$ is kept small, the Harmony and Prisoner's Dilemma games are not significantly affected, as shown in Figs.~\ref{fig:meanfield}(b,c). However, upon increasing the effective temperature $\theta$, the effective potential becomes flatter and smoother, rendering the system  more sensitive to the finite-size effects [see  Figs.~\ref{fig:meanfield}(e,f)]. 
In addition, the unstable area of the Stag Hunt game becomes wider while the region of mixed strategies in the Snowdrift game becomes narrower. More precisely, an increase of the effective temperature $\theta$ expands and deforms the $T-S$ diagram due to its unique dependence, at the mean-field level and for large $\mathcal N$, on the rescaled values of $t_{\theta}$ and $s_\theta$ (here $t_{\theta}=\frac{T-1}{T\theta}$ and $s_\theta=\frac{-S}{T\theta}$). Hence, a  temperature change from $\theta$ to $\theta'$ can be seen as the following mapping
\begin{equation}
  (T,S) \longrightarrow \left(\frac{\theta}{\theta'+(\theta-\theta')T}T,\frac{\theta'}{\theta'+(\theta-\theta')T}S\right).
\end{equation}
For instance, the point $(T,S,\theta)=(1,0.05,0.05)$ which has a fraction of cooperation of $1$ shifts to $(S',T',\theta')=(1,0.5,0.5)$; the point $(T,S,\theta)=(1.05,0.05,0.05)$ with a cooperation density of $1/2$ shifts to $(S',T',\theta')\simeq (1.9,0.9,0.5)$; and $(T,S,\theta)=(1.05,0,0.05)$ with a zero cooperation density moves to $(S',T',\theta')\simeq (1.9,0,0.5)$.

\begin{figure}[t!]
    \centering
    \includegraphics[width=0.45\textwidth]{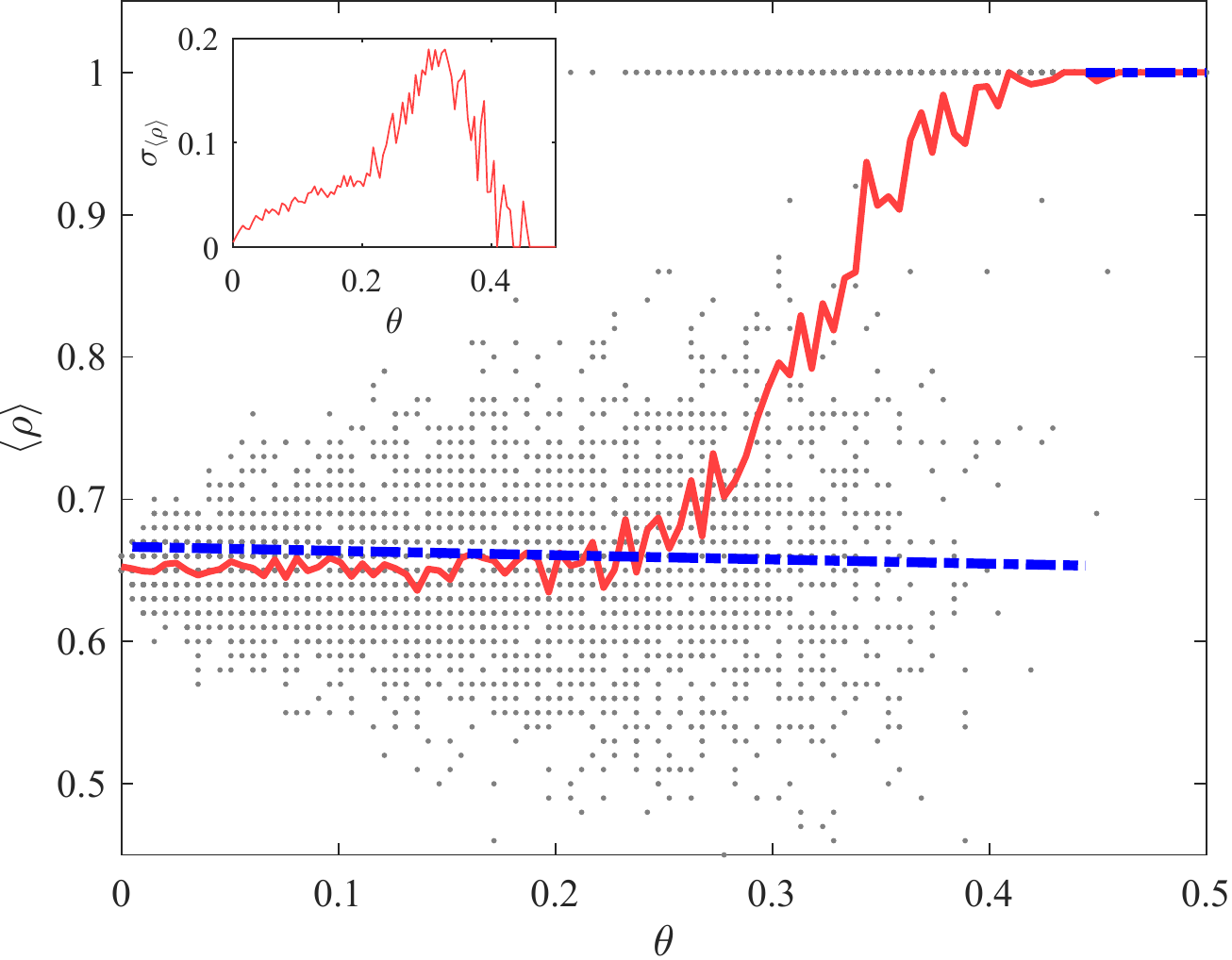}
    \caption{Average cooperation as a function of the effective temperature $\theta$ for a complete graph with $\mathcal N=100$ nodes. Each dot corresponds to different Monte Carlo simulations (50 in total) and the red curve is the sample average. Inset shows the standard deviation in the average cooperation. The game parameters are $T=1.2$ and $S=0.4$ (Snowdrift game). The blue dashed line for $\theta\le 0.45$ is the analytical value of the cooperation given by Eq.~\ref{eq:rhoc} when the effective potential \eqref{eq:prhost} has a local minimum $\rho\in (0,1)$, while for $\theta >0.45$, the effective potential has only one minimum located at $\rho_m=1$. }
    \label{fig:meanfield2}
 \end{figure}

Regarding the critical behavior of the system, the main difference between the cases $\theta=0$ and $\theta>0$ lies on the presence of discontinuous transitions. As a representative example, let us consider the point E in Fig.~\ref{fig:meanfield}(a) for $\theta=0$ to discuss how the fraction of cooperators evolves as the agents increase their stochastic behavior ($\theta$). Figure ~\ref{fig:meanfield2} shows the scatter plot of $\rho$ as the effective temperature increases for a game setting close to point E (each dot represents the outcome of a Monte Carlo simulation). While the sample average $\mean{\rho}$ (red curve) changes from $\mean{\rho}\sim 0.65$ to $\mean{\rho}\sim 1$ in a narrow region around $\theta=0.35$, the cloud of points spreads more as the temperature rises, up to a point near $\theta=0.4$ where the majority of the system outcomes is of full cooperation. The inset shows a peak in the cooperation fluctuations pointing towards a discontinuous temperature-induced transition. From the viewpoint of the effective potential, Figs.~\ref{fig:meanfield}(e,f), we observe how it loses its minimum at $\rho_m\simeq 0.65$ (dashed blue line given by Eq.~\eqref{eq:rhoc} for $\theta\le 0.045$) upon increasing $\theta$ [compare the magenta curves in panels (e) and (f)]. The system changes from $\rho_m\simeq 0.65$ for $\theta<\theta_c\simeq 0.35$ to a state of complete cooperation for $\theta>\theta_c$. The transition is discontinuous in the thermodynamic limit $\mathcal N \to \infty$. 

\subsection{Complex networks}

In order to explore the extent of our mean-field theory beyond the complete graph, we also consider random regular graphs. New interesting features emerge as it is confirmed by the Monte Carlo simulations in the panels (g), (h), and (i) of Fig.~\ref{fig:meanfield} for  random regular graphs with degree $k=3$, and the same values of the effective temperature as in the upper plots. We recall that the results are for the specific initial conditions of $\rho=0.5$ ($50\%$ of cooperators) and all agents randomly distributed all around the nodes of the network, regardless their the strategy. 

For zero effective temperature $\theta=0$, Fig.~\ref{fig:meanfield}(g), the $T-S$ plane divides into six clear disjoint domains with different cooperation densities. Within each domain, variations of the parameters $T$ and $S$ do not produce any change in the system's behavior, while crossing two adjacent domains induces discontinuous changes in the cooperation density as described in Sec.~\ref{sec:transitionpoints}. The dashed lines delimiting the different domains correspond to solutions to Eq.~\eqref{eq:diccond1}, showing an excellent agreement between theory and numerical simulations. Comparing panels (a) and (g), we notice how the  structured interactions favor the expansion of  full cooperation (region I) into adjacent games, while complete defection is limited now to a smaller region (VI).  
 The remaining four domains describe situations which can not be described using the mean-field framework. Three of them (regions III, IV, and V) are characterized by the presence of quasi-absorbing states of different cooperation densities, with region V invading three game quadrants but displaying a very low cooperation density with just a few cooperators and oscillating nodes. Finally, the region II in the Snowdrift quadrant exhibits mixed strategy states but, contrary to the all-to-all case, with only one possible intermediate level of cooperation.    

 When the effective temperature is slightly increased up to $\theta=0.05$, see Fig.~\ref{fig:meanfield}(h), the distribution of the different dynamical regimes observed for $\theta=0$ keeps more or less similar, but the sharp boundaries become smoother. As already reported in Fig.~\ref{fig:noise}(a,b), the cooperation density changes continuously with the temptation-to-defect parameter, as soon as the choice to change strategy is no longer deterministic. 
This effective temperature, although small, is enough to destroy the quasi-absorbing states in region V, enlarging the domain with pure defection. However, the mixed strategy states of region II and the quasi-absorbing states of regions III and IV stay almost with the same levels of average cooperation, except for a not negligible range of parameter settings where cooperation is promoted. A further increase of the effective temperature $\theta$, up to $0.5$ in Fig.~\ref{fig:meanfield}(i), completely ``destroys'' all the quasi-absorbing states, in the sense that the system keeps away from them. 
As already discussed when describing  Figs.~\ref{fig:noise}(e,f), this scenario is consistent with a discontinuous transition with the effective temperature as a control parameter. 

\section{Conclusion}
\label{sec:conclusion}

We have studied the dynamics of cooperators and defectors on an structured environment when playing different cooperative games subject to eventual irrational choices. Overall, the system exhibits emergent complex behavior, which includes abrupt and continuous transitions as we change the probability of the irrational choices (tuning the effective temperature $\theta$), the parameters of the game (entries of the generic payoff matrix), and the structure of the interactions through the topology of the underlying network.

For finite system size, we have identified the most general steady states of the system, given in terms of the absorbing (consensus), quasi-absorbing, and mixture strategy states (which all form absorbing sets of states). Moreover, we have also obtained necessary and sufficient conditions for the existence of discontinuous transitions when $\theta=0$ (deterministic interactions). They include a geometric condition, given by Eq.~\eqref{eq:diccond1}, which involves the parameters of the payoff matrix and the degrees sequence of the interaction network; and a dynamic condition that requires the existence of agents in the geometric condition. It has also been shown that for $\theta>0$ the previous transitions are 
continuous. 

In the simplest interaction scenario, when all agents interact with all others, the system can be completely described by the fraction of cooperators $\rho$. An exact master equation for the probability of $\rho$ has been used to derive a more tractable Fokker-Planck equation, suitable for describing the system for (typical) large system sizes. Then, a regularized solution to the Fokker-Planck equation, after removing the divergences induced by the absorbing states, has been obtained. This solution provides an explicit expression for an effective potential which describes correctly the behavior of the system not too close to the absorbing (consensus) states. This has allowed us to explicitly assess the finite-size effects and the impact of the effective temperature on the dynamics. We recover the replicator equation for large system sizes and small effective temperatures: the effective potential has a local minimum describing a coexistence of strategies only in the parameter region of the Snowdrift game ($0<S<1<T<2$). Due to a scaling property of the effective potential, we have also seen that increasing the effective temperature $\theta$ is equivalent to keeping it constant and changing the game parameters properly. In particular, upon increasing $\theta$, the coexistence region shrinks and moves to higher values of $T$ and $S$. Interestingly, in large systems, increasing the effective temperature may induce a discontinuous transition in the level of cooperation: the local minimum of the effective potential around $\rho_m\in(0,1)$ disappears above a critical value of $\theta$ and only the minima describing consensus survive.   

The previous mean-field scenario becomes more complex when the network of interactions are structured. First, we observe discontinuous transitions for zero effective temperature not explained by mean field. 
Moreover, the game parameter space splits into domains of differentiated dynamical regimes, separated by discontinuous transitions between them. The transitions are ruled by the condition \eqref{eq:diccond1} with an additional dynamical condition, which, eventually, makes the domains depend on the initial conditions. That is, different initial percentages of cooperators may select another set of transitions among the solutions to the Eq.\eqref{eq:diccond1}. Second, while in most of the games the qualitative behavior of the system is well captured by mean-field, the presence of quasi-absorbing states is a new and interesting ingredient. What the numerical simulations show is the presence of connected domains of nodes with frozen strategy, separated by a frontier of frustrated agents with oscillating strategy. This gives rise to an intermediate level of cooperation mostly located in the region of parameters of the Stag Hunt game, but also for the Snowdrift and Prisoner's Dilemma games. As a consequence, this suggests the existence of an effective potential, a function of the fraction of cooperation, that develops a minimum in this region of the parameter space, unlike the mean-field case. Our simulations also show that upon increasing the effective temperature above a critical value (which is a nontrivial function of the parameters of the system), the system abruptly moves away from the quasi-absorbing states, which can be interpreted as a change in the local minimum of the potential to a maximum, hence recovering some of the predictions of mean-field. Finally, the system exhibits interesting non-monotonous phenomena, which is a new feature not present in the mean-field theory. This happens in the two games showing coexistence of opinions, the Snowdrift and the Stag Hunt games, but with different peculiarities. In the Snowdrift game, there is a region of parameters where the cooperation density develops a minimum as a function of the effective temperature, a surprising instance of stochastic resonance. In the Stag Hunt game, we observe a similar behavior when the effective temperature is above the critical one and the quasi-absorbing states are destroyed. Now, the probability of reaching full cooperation exhibits a maximum for intermediate values of $\theta$. 

In conclusion, the evolution of cooperation in evolutionary social systems is critically determined by the underlying structure of interactions among agents and their level of irrational choices. Here, we have provided a deeper insight into the network reciprocity mechanism by describing abrupt shifts in the cooperation, due to particular arrangements of the network interactions and purely deterministic strategy updates, as well as genuine phase transitions and stochastic resonances induced by an effective temperature calibrating the stochastic nature of the social behavior.

\section*{Acknowledgments}
This research was supported by the Spanish  Ministerio de Ciencia e Innovación (Project PID200-113737GB-I00), by Rey Juan Carlos University (Grant M2605), and by Community of Madrid and Rey
Juan Carlos University through Young Researchers program in R\&D (Grant CCASSE M2737).  

\appendix

\section{The case of zero effective temperature}
\label{app:A}
For $\theta=0$ the copying mechanism is (almost) deterministic. Taking the limit $\theta\to 0$ in Eq.~\eqref{eq:copprob} we get
\begin{equation}
  \label{eq:psntz}
  p_{\sigma,\nu}=\Theta (g_\nu-g_\sigma)
\end{equation}
for the probability of the selected node $\sigma$ to copy the selected neighbor $\nu$. The $\Theta$ function only let the probability have three values: zero ($g_\nu<g_\sigma$), one ($g_\nu>g_\sigma$) or one half ($g_\nu=g_\sigma$). 

For the case of all-to-all interactions, the relevant quantities are the probabilities $p^\pm$ of increasing ($+$) and decreasing ($-$) the number of cooperators  by one. They can be written as
\begin{equation}
p^\pm =\Theta\left[ \mp (T-R+S-P)\rho \mp \left(\frac{\mathcal N-1}{\mathcal N}P+\frac{R}{\mathcal N}-S\right) \right].    
  \nonumber 
\end{equation}
  
The arguments of the Theta function are linear functions of $\rho$ that cancel out for $\rho$ equal to
\begin{equation}
  \rho_m^\pm=\frac{S-\frac{\mathcal N-1}{\mathcal N}P-\frac{R}{\mathcal N}}{T-R+S-P}.
\end{equation}

The behavior of the system is determined by the two limiting conditions $\rho_m=\frac{1}{\mathcal N},1-\frac{1}{\mathcal N}$. Note that we do not consider $\rho_m=0,1$ since the previous possible values of $\rho$ already may induce the system to reach the consensus states surely. For the typical values $R=1$ and $P=0$, the previous conditions give
\begin{eqnarray}
  && \rho_m=\frac{1}{\mathcal N}\quad \Rightarrow \quad S=\frac{1}{\mathcal N-1}+\frac{T-1}{\mathcal N-1}, \\
  && \rho_m=\frac{\mathcal N-1}{\mathcal N}\quad \Rightarrow \quad S=1+(\mathcal N-1)(T-1),
\end{eqnarray}
both representing lines in the $T-S$ plane with small and large slope, respectively.

\section{Some properties of the effective potential}
\label{appen:1}

\subsection{Limit of zero effective temperature}
In the limit of zero effective temperature, $\theta \to 0$, Eq.~\eqref{eq:veff} for the effective potential reduces to
\begin{equation}
V(\rho)=\frac{1}{t_\theta-s_\theta}|t_\theta\rho+s_\theta(1-\rho)|+\frac{1}{2\mathcal N}\ln[\rho(1-\rho)+\kappa],
\end{equation}
which is independent of $\theta$. If we remove the finite-size contribution, the potential has an extreme at $\rho=\rho_0$, given by Eq.~\eqref{eq:rho0}. It is a minimum when $t_\theta>s_\theta$, and a maximum otherwise. 

\subsection{Local extremes}

From Eq.~\eqref{eq:prhost} we see that, in most of the cases, the width of the distribution $\mathcal P$ around a maximum is of the order of $1/\mathcal N$, hence the relevant contributions to the distribution are located at the minima of the effective potential. Apart from the absorbing states, $\rho=0,1$, the effective potential has an additional minimum at $\rho_m\in(0,1)$, when $V'(\rho_m)=0$ and $V''(\rho_m)>0$.

The derivative of the effective potential is
\begin{equation}
  \label{eq:vrhop}
V'(\rho)=\tanh\left[\frac{t_\theta\rho+s_\theta(1-\rho)}{2}\right]+{\frac{1}{2\mathcal N}}\frac{1-2\rho}{\rho(1-\rho)+\kappa},
\end{equation}
even if $t_\theta=s_\theta$. It can be graphically seen that the equation $V'(\rho_m)=0$ has three solutions at most, for any values of $t_\theta,\, s_\theta,\, \mathcal N$, and $\kappa$. When $\theta\ll\mathcal N$, in the regime (a) discussed in Sec.~\ref{sec:meanfield}, we can localize an extreme of $V(\rho)$ at $\rho_m$ given by Eq.~\eqref{eq:rhoc}. With the same accuracy, we can also compute the second derivative, with the result given by Eq.~\eqref{eq:vpp}.

We can also obtain exact results, useful for understanding the behavior of the system as parameters are changed. Consider the case of $s_\theta=-t_\theta$ (equivalently, $P-S=R-T$). The drift and diffusion terms have the same symmetry, they are odd functions of $1-\rho/2$. Hence, an exact solution to the equation $V'=0$ is
\begin{equation}
  \rho_m=\frac{1}{2}. 
\end{equation}
Moreover, the second derivative of the potential at $1/2$ is $t_\theta-\frac{4}{\mathcal N(1+4\kappa)}\simeq t_\theta-\frac{4}{N}$, meaning that $V$ has a local minimum at $\rho_m$ when
\begin{equation}
  -s_\theta=t_\theta> \frac{4}{\mathcal N(1+4\kappa)}\simeq \frac{4}{\mathcal N}. 
\end{equation}
Note that the previous condition is not restricted to any value of the system size nor the effective temperature. For $\mathcal N\to \infty$ we recover condition $t_\theta-t_s>0$, valid for regime (a). Moreover, for $R=1$ and $P=0$, the previous condition reads $S=T-1> \frac{4\theta}{\mathcal N}T$, which gives the critical condition $T> \frac{1}{1-\frac{4\theta}{\mathcal N}}$.

For $t_\theta=s_\theta$, both the diffusion and drift terms have the same symmetry as well, since the drift term does not depend on $\rho$. It is readily seen that in this case the equation $V'=0$ has two solutions, only one solution ($\rho_m$) being in $(0,1)$ for any values of the parameters. However, it turns out that $V''(\rho_m)>0$, meaning that when $T-R=P-S$ the system always ends up at an absorbing state.  

\bibliography{references}

\end{document}